\documentclass[12pt,thmsa]{article}
\usepackage{sw20lart}

\input{tcilatex}
\begin{document}

\author{{\small Ximing Fang, Xiwen Zhu, Mang Feng, Xi'an Mao and Fei Du} \\
%EndAName
{\small Laboratory of Magnetic Resonance and Atomic and Molecular Physics, }%
\\
{\small Wuhan Institute of Physics and Mathematics, The Chinese Academy of
Sciences, }\\
{\small Wuhan 430071, P. R. China}}
\title{Experimental Implementaton of Dense Coding Using Nuclear Magnetic Resonance}
\maketitle

\begin{abstract}
Quantum dense coding has been demonstrated experimentally in terms of
quantum logic gates and circuits in quantum computation and NMR technique.
Two bits of information have been transmitted through manipulating one of
the maximally entangled two-state quantum pair, which is completely
consistent with the original ideal of the Bennett-Wiesner proposal. Although
information transmission happens between spins over inter-atomic distance,
the scheme of entanglement transformation and measurement can be used in
other processes of quantum information and quantum computing.


\end{abstract}

\begin{center}
Introduction
\end{center}

While the miraculous properties of entangled states of quantum systems are
revealed and understood, more attention is paid to the use of such entangled
states for quantum information transmission and processing. Among the major
appplication schemes investigated are cryptography based on Bell's theorem,
teleportation and dense coding. As sending more than one bit of information
classically requires manipulation of more than two-state particle, dense
coding proposed by Bennett and Wiesner\cite{a}in 1992, which can transmit
two bits message by manipulating one of the entangled two-state pair, is
seemingly striking. Several alternatives for realizing dense coding have
been suggested \cite{b,c,d,e,f,g}. Quite recently, Shimizu et al\cite{h}%
proposed a scheme, with which the capacity can be enhanced to 3 bits by the
local operation of a frequency-dependent phase shift on one of polarization
entangled twin photons. However, only one experiment of dense coding has
been reported\cite{e}so far. In Ref.\cite{e}, the polarization-entangled
photon pair was exploited for transmitting 3 messages per two-state photon,
i. e. 1 'trit'$\approx $1.58bit.

One of the reasons why the quantum computer can outperform the classical
counterpart is that mutil-particle entanglement is widely used for
information processing in quantum computing. It is not surprising that there
is much in common between quantum computation and dense coding as the latter
is actually a process of generation, manipulation and measurement of
entangled states for two-particle systems.Many schemes for realizing quantum
computation have been proposed, some of them, e. g. trapped ultra-cold ions%
\cite{i},cavity-QED\cite{j}and nuclear magnetic resonance(NMR)\cite{k,l},
have been experimentally demonstrated for a single quantum logic gate.
However, the real experiments\cite{m,n,o,p}of executing a quantum algorithm
have been solely performed with NMR due to its physical advantages and
technical maturity. While the ensemble averaging over nuclear spin-$\frac{1}{%
2}$molecules in a bulk solution sample has been chosen as a two-state
quantum system, its quantum behavior is equivalent to that of any real
two-state micro-particle. Furthermore, principles and techniques developed
are able to be used for constructing the logic gate and quantum circuits in
quantum computation conveniently. As an interesting example of quantum
information transmission, quantum teleportation was proposed to achieve in
terms of primitive operations\cite{q}in quantum computation and
experimentally demonstrated\cite{r}by using technique of NMR quantum
computing with molecules of trichloroethylene over inter-atomic distances.
Therefore, it is natural to attempt to do another type of quantum
communication, dense coding, in much the same way. Here we report our
experiment of quantum dense coding with the help of \thinspace logic gates
in quantum computation and NMR technique, trransmiting 4 messages ,i. e. 2
bits, by treating one of the entangled quantum pair, exactly as the original
meaning of the Bennett-Wiesner proposal, over angstroms distance.

The concept of dense coding is briefly discussed here(see Fig. 1). Alice,
the information reciever, prepares a pair of Einstein-Podolsky-Rosen (EPR)
particles and sends one of them to Bob, the information sender, while
retaining the other at her hand. When Bob wants to communicate with Alice,
he can encode his message by manipulating the particle sent by Alice with
one of the four transformation of \thinspace I,$\sigma _{z},\sigma _{x\text{ 
}}$and $i\sigma _{y}$, where $\sigma _{x\text{ }},$ $\sigma _{y}$and$\sigma
_{z}$are the Pauli matrices and I is the unit matrix, then transmit the
encoded particle to Alice. Alice decodes her recieved message by measuring
the two particles at her hand and learns which transformation Bob applied.

\section{The network of quantum dense coding}

The network that we use for realizing quantum dense coding in terms of the
logc gates and circuits in quantum computation is shown in Fig. 2, where a
and b denote two quantum systems with two states \TEXTsymbol{\vert}0%
\TEXTsymbol{>} and \TEXTsymbol{\vert}1\TEXTsymbol{>}. It consists of three
parts separated by the dotted lines. Our strategy includes three steps.

\subsection{Preparation of the entangled state (EPR pair)}

The EPR pair $\frac{1}{\sqrt{2}}\left( |00>-|11>\right) $ can be prepared by
the operations of the NOT gate N$_{b}$, the Walsh-Hardamard gate H$_{b}$and
the controlled-NOT (C-NOT) gate CN$_{ba}$. Supposing that the input state is 
\TEXTsymbol{\vert}0\TEXTsymbol{>}$_{b}$\TEXTsymbol{\vert}0\TEXTsymbol{>}$_{a}
$, we have

\begin{equation}
\begin{array}{l}
|0>_{b}|0>_{a}\underrightarrow{N_{b}}|1>_{b}|0>_{a} \\ 
\underrightarrow{H_{b}}\frac{1}{\sqrt{2}}\left( |0>-|1>\right) _{b}|0>_{a}
\\ 
\underrightarrow{CN_{ba}}\frac{1}{\sqrt{2}}\left( |00>-|11>\right)
\end{array}
\end{equation}
It is easily seen that the other three Bell base states $\frac{1}{\sqrt{2}}%
\left( |00>+|11>\right) ,$ $\frac{1}{\sqrt{2}}\left( |01>-|10>\right) $ and $%
\frac{1}{\sqrt{2}}\left( |01>+|10>\right) $ can be produced by replacing N$%
_{b}$ with I (the identity transformation, i. e. nothing being done), N$_{b}$%
N$_{a}$ and N$_{a},$ respectively.

\subsection{Transformation of the entangled states}

Through operating one member of the EPR pair with four different
transformation U$_{ai}$=I, $\sigma _{z}$, $\sigma _{x}$ and $i\sigma _{y}$
(for i=1, 2, 3 and 4), one obtains four Bell base states respectively, thus
achieving message encoding. The processes are represented as

\begin{equation}
\frac{1}{\sqrt{2}}\left( |00>-|11>\right) \rightarrow 
\begin{array}{l}
\underrightarrow{U_{a1}}\frac{1}{\sqrt{2}}\left( |00>-|11>\right) \\ 
\underrightarrow{U_{a1}}\frac{1}{\sqrt{2}}\left( |00>+|11>\right) \\ 
\underrightarrow{U_{a1}}\frac{1}{\sqrt{2}}\left( |01>-|10>\right) \\ 
U_{a1}-\frac{1}{\sqrt{2}}\left( |01>+|10>\right)
\end{array}
\end{equation}

\subsection{Measurement of the entangled states}

It is accomplished by the action of the C-NOT gate CN$_{ba}$ and the gate H$%
_{b}$. After these measurements that can distinguish different Bell base
states, four states, $|y_{bi}>|x_{ai}>\equiv |yx>_{i}$=\TEXTsymbol{\vert}10%
\TEXTsymbol{>}, \TEXTsymbol{\vert}00\TEXTsymbol{>}, \TEXTsymbol{\vert}11%
\TEXTsymbol{>} and -\TEXTsymbol{\vert}01\TEXTsymbol{>} for i = 1, 2, 3 and 4
respectively are read out and the messages are decoded.

It can be clearly seen that four messages have been transmitted from Bob to
Alice via encoding transformation on one member of the entangled pair and
decoding measurement of the whole system in terms of the network shown in
Fig. 2, which is consisttent with the original idea of the Bennett-Wiesner
proposal. If quantum system a (e. g. photon) could move far away from b in
the processes from step 1 to step 2 and step 2 to step 3, which means Alice
locating a long distance away from Bob, then one fulfills long-distance
communication. If a (e. g. atom or nuclear spin) would stay in the vicinity
of b in the whole process, which means that Alice and Bob are close to each
other, then close communication is brought about in this case. No matter
which of the four Bell states the process begins from, we always get
definite outputs associated with different transformation operations. The
relationship between the encoding transformation U$_{ai}$and the decoding
outputs $|yx>_{i}$under various starting EPR states is listed in table 1.

\section{The experimental procedure and result}

The network stated above can be put into practice with NMR techniques. We
chose $^{1}H$ and $^{13}C$ in the molecule of carbon-13 labeled choroform $%
^{13}CHCl_{3}$ (Cambridge Isotope Laboratories,Inc.) as the two-spin system
in the experiments and d6-acetone as the solvent, the solute/solvent ration
being 1:1(v/v).The sample was flame sealed in a standard 5-mm NMR tube.
Spectra were recorded on a Bruker ARX500 spectrometer with a probe tuned at
500.13MHz for $^{1}H$ (denoted by a), and 125.77MHz for $^{13}C$ (denoted by
b).

The NOT gate was realized by applying a pulse of $X\left( \pi \right) $, H$%
_{b}$ gate was implemented by using pulses of $X\left( \pi \right) Y\left( -%
\frac{\pi }{2}\right) $, and the C-NOT gate was realized by the pulse
sequence shown in Fig. 3. Each pair of $X\left( \pi \right) $ pulses applied
on the spin took opposed phases in order to reduce the error accumulation
caused by imperfect calibration of the $\pi $ pulses. The $U_{ai}s$
transformations correspond to pulses through $U_{a1}\sim I$, $U_{a2}\sim
Z_{a}\left( \frac{\pi }{2}\right) $, $U_{a3}\sim X_{a}\left( \pi \right) $
and $U_{a4}\sim Y_{a}\left( \pi \right) $.

The experiments were carried out in a procedure as follows. Firstly, the
effective pure state \TEXTsymbol{\vert}00\TEXTsymbol{>} was prepared by
temporal averaging\cite{m}. Then the operations applied according to the
network shown in Fig. 2 were executed by a series of NMR pulses mentioned
above. Finally, the outputs of experiments, described by the density
matrices $\rho _{outi}=|yx>_{ii}<yx|$ corresponding to four operations of $%
U_{ai},$were reconstructed by the technique of state tomography\cite{m}.

The matrices reconstructed by fitting the measured data  from the recorded
spectra  are shown in Fig. 4. The axes of the horizontal plane denote the
locations of the elements in the matrix, its values 0, 1, 2 and 3 correspond
to the states \TEXTsymbol{\vert}00\TEXTsymbol{>}, \TEXTsymbol{\vert}01%
\TEXTsymbol{>}, \TEXTsymbol{\vert}10\TEXTsymbol{>} and \TEXTsymbol{\vert}11%
\TEXTsymbol{>}, respectively, and the vertical axis represents the module of
the theoretical and experimental matrices, we can find that they are in good
agreement. Experimental errors were primarily due to the inhomogeneity of
the RF field and static magnetic field, inaccurate calibration of pulses and
signal decay during the experiment. The largest error is about 10\%. These
results show that the network actually does the quantum dense coding
communication.

\section{Conclusion}

We have experimentally implemented quantum dense coding by using NMR quantum
logic and circuits in quantum computation. With the help of non-classical
effects inherent in quantum systems, four messages, i.e., 2 bits of
information (instead of three messages or 1.58 bit as in Ref.\cite{e}) have
been transmitted via treating one of the entangled two spin-$\frac{1}{2}$
systems, which demonstrated physically the original idea of the
Bennett-Wiesner proposal. Principlly, the dense coding experiment with
maxially entangled states for n\TEXTsymbol{>}2 particles can be performed by
utilizing an appropriate NMR sample with n spins and a generalized quantum
network. As the nuclear spins in a molecule are coupled through chemical
bond, the communication by dense coding via NMR technique actually conducts
between spins in angstrom distance. Nevertheless, the concept and the method
of transformation and measurement of the maximally entangled states in terms
of the quantum computation network should be useful in other processesof
quantum information processing and quantum computing.

\subsection{Acknowledgements}

This work was supported by National Natural Science Foundation of China and
Chinese Academy of Sciences.

*To whom correspondece should be addressed. E-mail address:
xwzhu@nmr.whcnc.ac.cn

\section{\newpage}

Captions of the Figures and Table

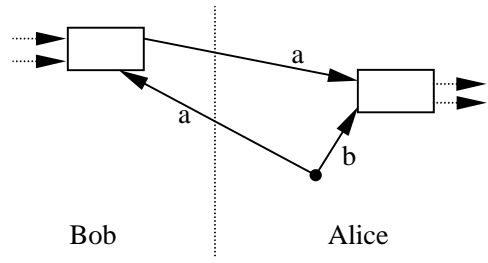
Fig. 1. The schematic diagram of dense coding.

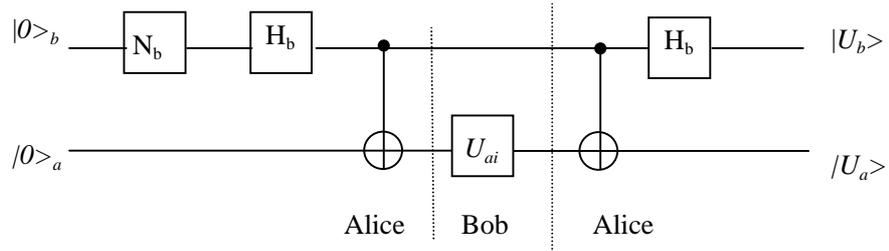
Fig. 2. The network for quantum dense coding.

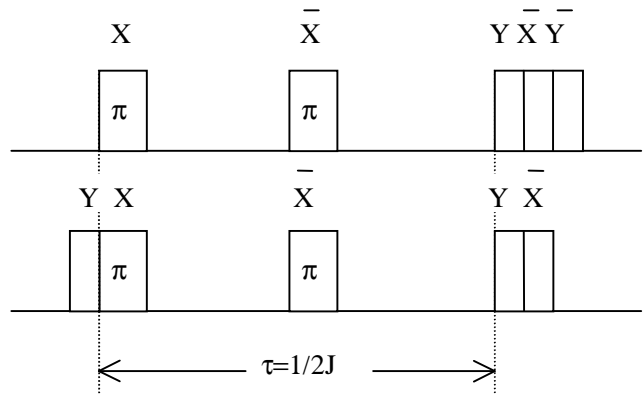
Fig. 3. The pulses of controlled-NOT gate.

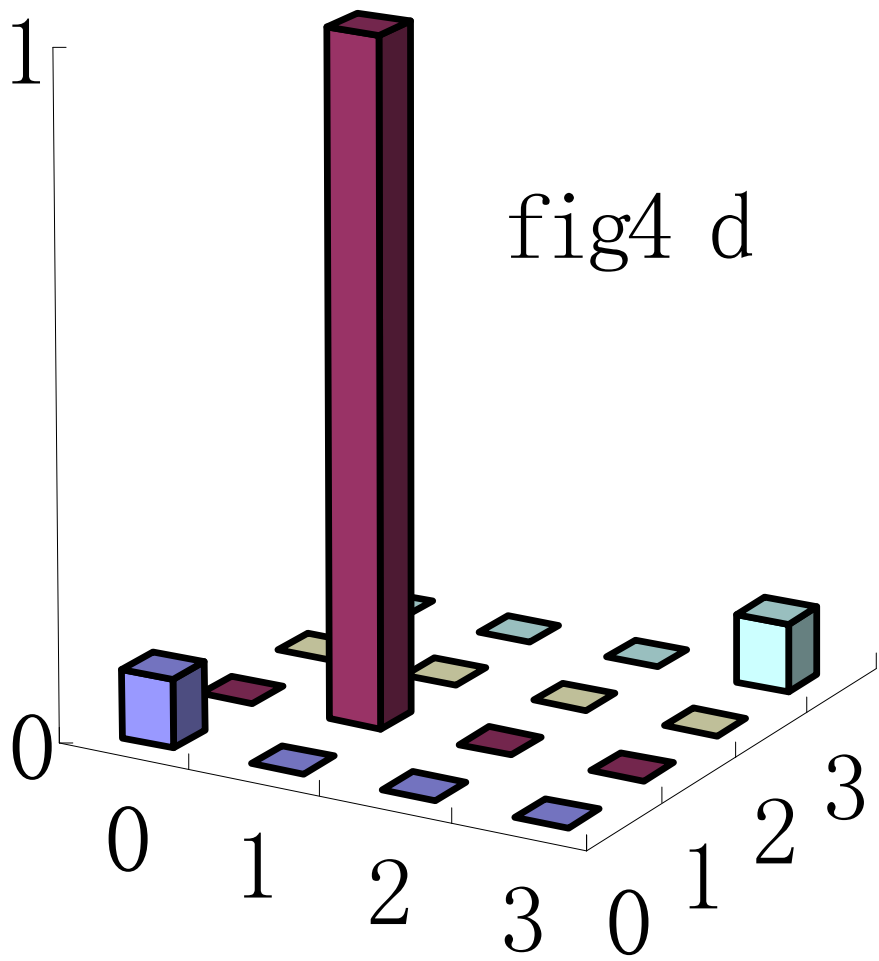
Fig. 4. The distribution of  the density matrices of the quantum system. The
a, b, c and d are the experimental results,  and e, f, g and h are the
theoretical ones corresponding \thinspace to to the operations I, $\sigma
_{z}$, $\sigma _{x}$ and $i\sigma _{y}$ respectively.

Table 1 Correspondence between the starting states and output states.

\newpage

Table 1

\begin{tabular}{lllll}
& $\frac{1}{\sqrt{2}}\left( |00>-|11>\right) $ & $\frac{1}{\sqrt{2}}\left(
|00>-|11>\right) $ & $\frac{1}{\sqrt{2}}\left( |00>-|11>\right) $ & $\frac{1%
}{\sqrt{2}}\left( |00>-|11>\right) $ \\ 
U$_{a1}$ & \TEXTsymbol{\vert}10\TEXTsymbol{>} & \TEXTsymbol{\vert}00%
\TEXTsymbol{>} & \TEXTsymbol{\vert}11\TEXTsymbol{>} & \TEXTsymbol{\vert}01%
\TEXTsymbol{>} \\ 
U$_{a2}$ & \TEXTsymbol{\vert}00\TEXTsymbol{>} & \TEXTsymbol{\vert}10%
\TEXTsymbol{>} & -\TEXTsymbol{\vert}01\TEXTsymbol{>} & -\TEXTsymbol{\vert}11%
\TEXTsymbol{>} \\ 
U$_{a3}$ & \TEXTsymbol{\vert}11\TEXTsymbol{>} & \TEXTsymbol{\vert}01%
\TEXTsymbol{>} & \TEXTsymbol{\vert}10\TEXTsymbol{>} & \TEXTsymbol{\vert}00%
\TEXTsymbol{>} \\ 
U$_{a4}$ & \TEXTsymbol{\vert}01\TEXTsymbol{>} & -\TEXTsymbol{\vert}11%
\TEXTsymbol{>} & \TEXTsymbol{\vert}00\TEXTsymbol{>} & \TEXTsymbol{\vert}10%
\TEXTsymbol{>}
\end{tabular}

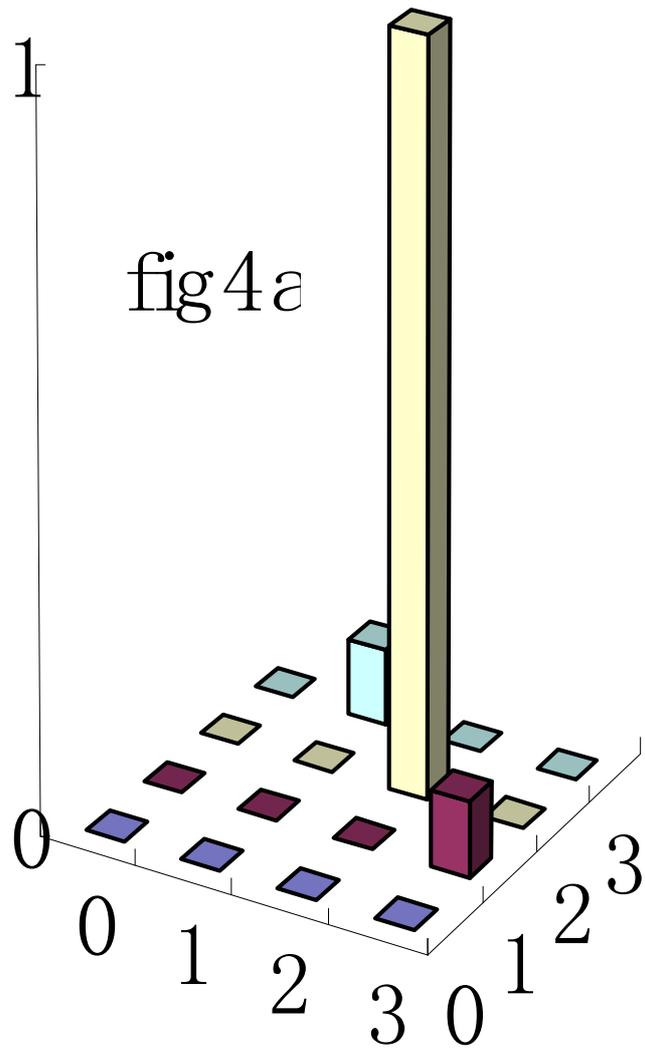
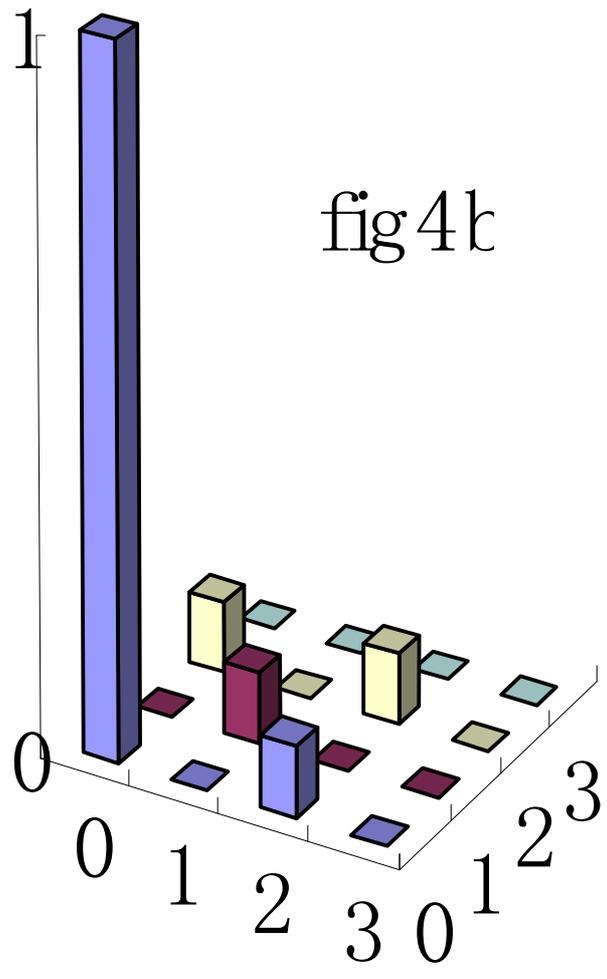
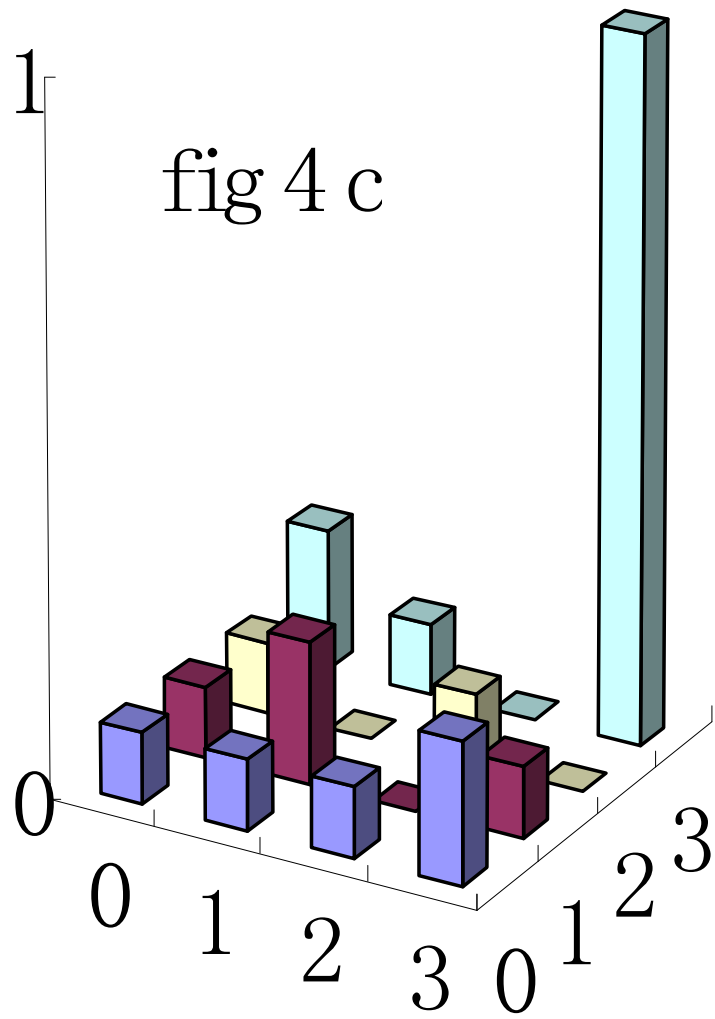
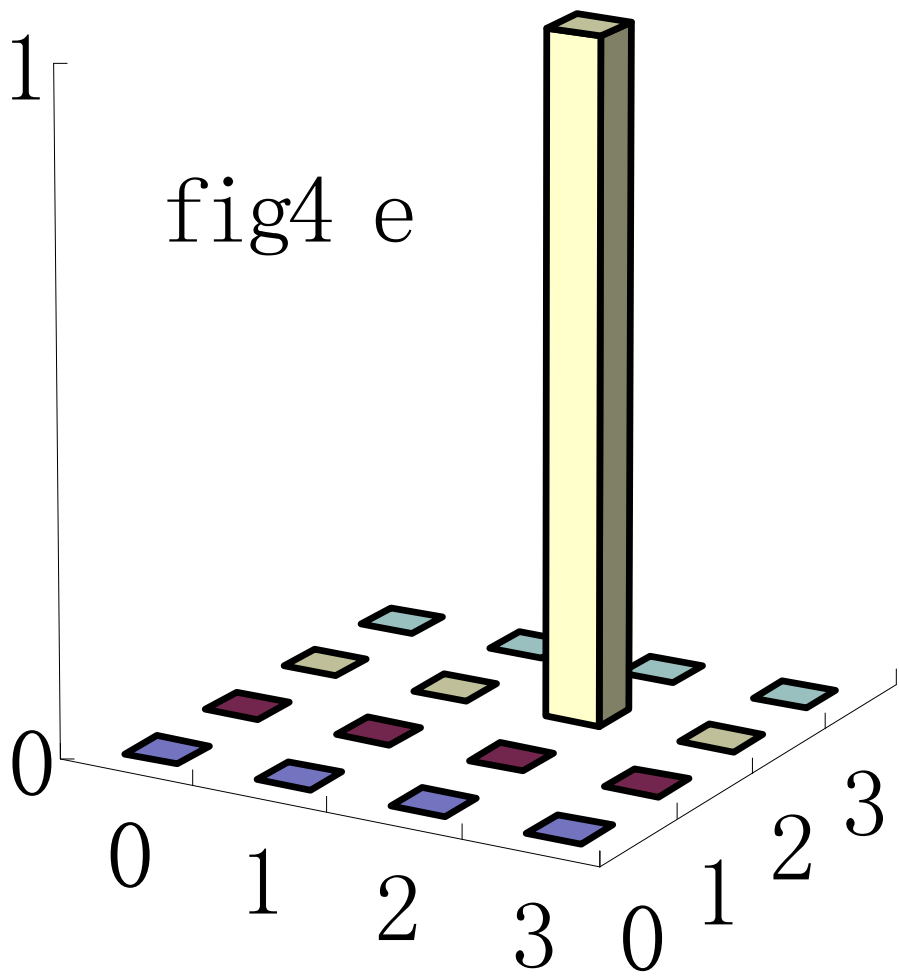
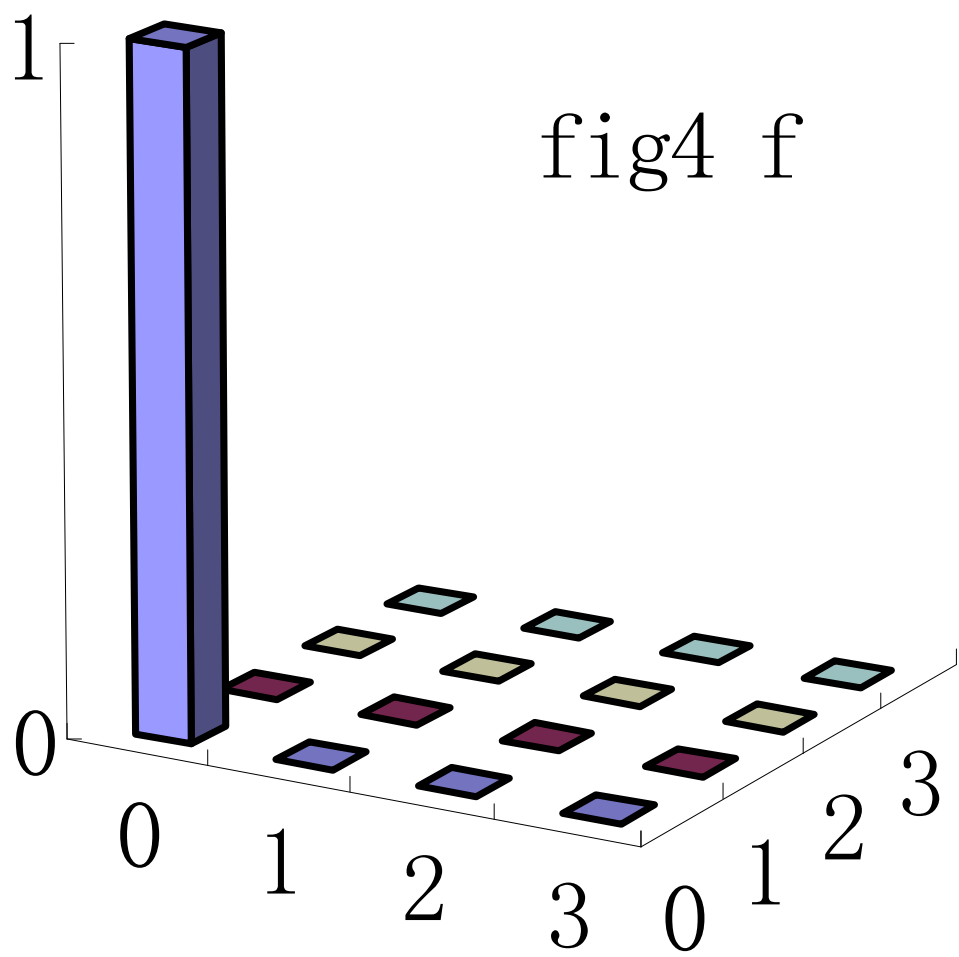
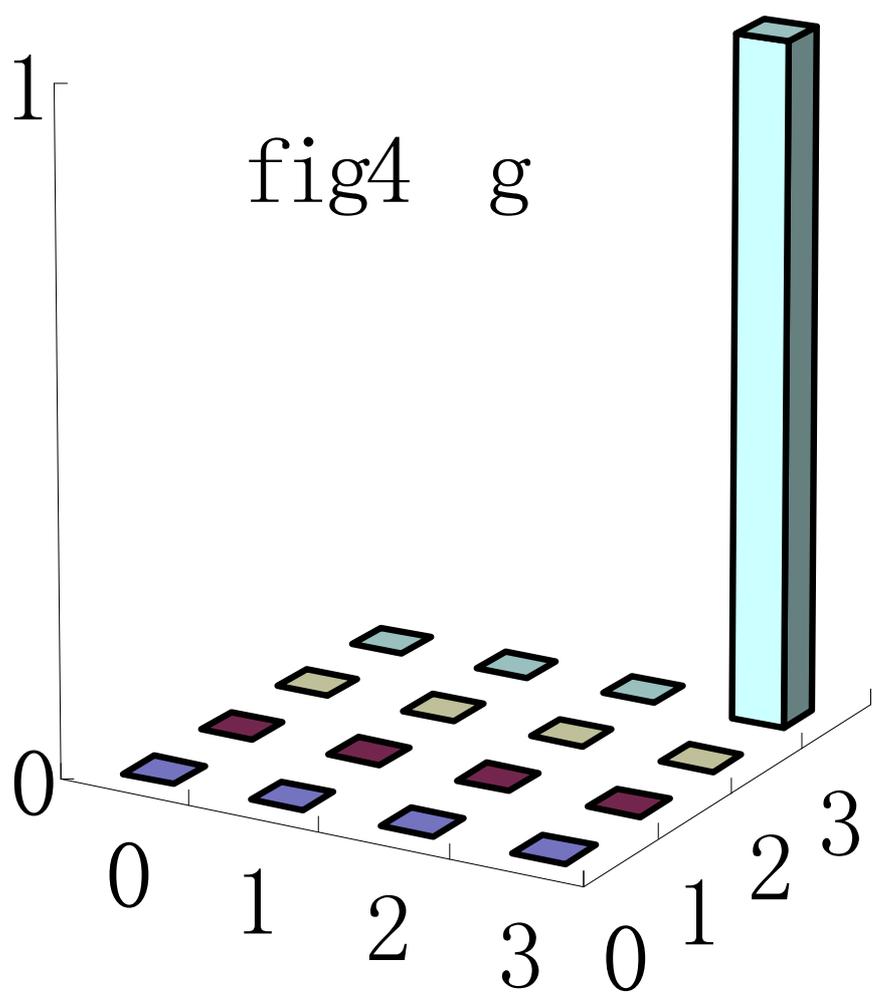
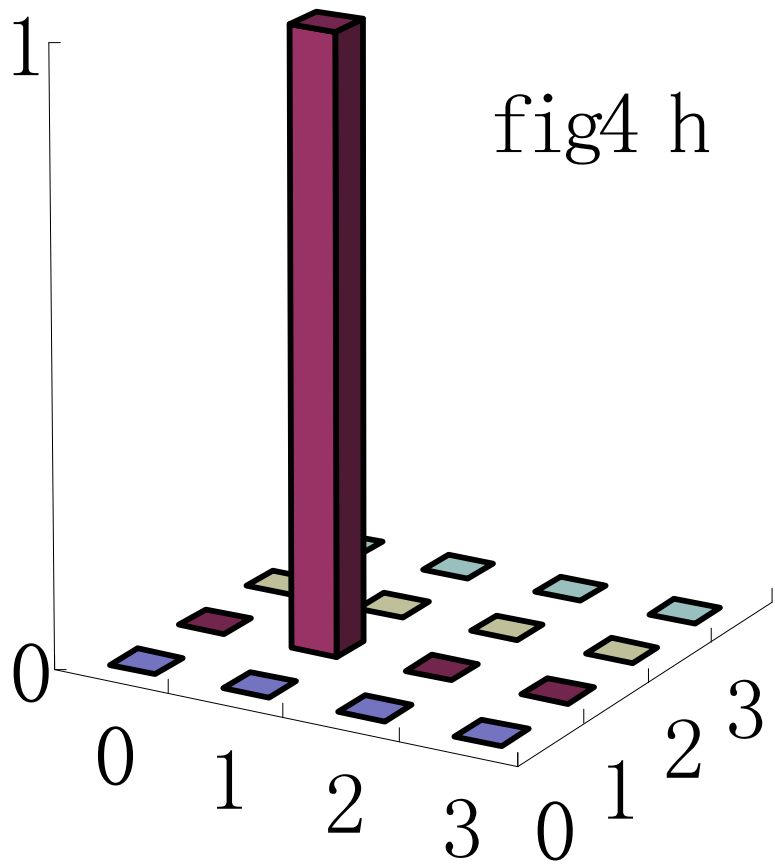
\end{document}